\documentclass[10pt,aps,prb,twocolumn, nofootinbib]{revtex4-1}

\usepackage{amssymb,amsmath}
\usepackage{graphicx}
\usepackage{color}
\usepackage{hyperref}
\usepackage{listings}
\usepackage{natbib}

\graphicspath{{figures/}}

\begin{document}

\title{\Large A Microphotonic Astrocomb}
\author{E. Obrzud$^{1,2}$, M. Rainer$^{3}$, A. Harutyunyan$^{4}$, M.H. Anderson$^{5}$, M. Geiselmann$^{5,6}$, B. Chazelas$^{2}$, S. Kundermann$^{1}$, 
S. Lecomte$^{1}$, M. Cecconi$^{4}$, A. Ghedina$^{4}$, E. Molinari$^{4,7}$, F. Pepe$^{2}$, F. Wildi$^{2}$, F. Bouchy$^{2}$, T.J. Kippenberg$^{5}$, T. Herr$^{1,*}$}

\affiliation{\footnotesize{\mbox{$^{1}$Swiss Center for Electronics and Microtechnology (CSEM), Time and frequency, Rue de l'Observatoire 58, 2002 Neuch\^atel, Switzerland}\\\mbox{$^{2}$Geneva Observatory/PlanetS, Department of Astronomy, University of Geneva, Chemin des Maillettes 51, 1290 Versoix, Switzerland}\\\mbox{$^{3}$National Institute of Astrophysics (INAF), Astronomical Observatory of Brera, Via Brera 28, 20121 Milano, Italy}\\\mbox{$^{4}$Fundaci\'on Galileo Galilei - INAF, Rambla Jos\'e Ana Fern\'andez P\'erez 7, 38712 Bre\~na Baja, Santa Cruz de Tenerife, Spain}\\\mbox{$^{5}$ Swiss Federal Institute of Technology (EPFL), SB IPHYS LPQM1, PH D3, Station 3, 1015 Lausanne, Switzerland}\\\mbox{$^{6}$ Ligentec, EPFL Innovation Park, B\^atiment C, 1015 Lausanne, Switzerland}\\\mbox{$^{7}$ INAF - Osservatorio Astronomico di Cagliari, Via della Scienza 5 - 09047 Selargius (CA), Italy}}\\$^*$tobias.herr@csem.ch}

\begin{abstract}
One of the essential prerequisites for detection of Earth-like extra-solar planets or direct measurements of the cosmological expansion is the accurate and precise wavelength calibration of astronomical spectrometers. It has already been realized that the large number of exactly known optical frequencies provided by laser frequency combs (“astrocombs”) can significantly surpass conventionally used hollow-cathode lamps as calibration light sources. A remaining challenge, however, is generation of frequency combs with lines resolvable by astronomical spectrometers. Here we demonstrate an astrocomb generated via soliton formation in an on-chip microphotonic resonator (“microresonator”) with a resolvable line spacing of 23.7 GHz. This comb is providing wavelength calibration on the 10 cm/s radial velocity level on the GIANO-B high-resolution near-infrared spectrometer. As such, microresonator frequency combs have the potential of providing broadband wavelength calibration for the next-generation of astronomical instruments in planet-hunting and cosmological research. 
\end{abstract}

\date{\today}

\maketitle

The existence of life on other planets and the evolution of our Universe are questions that extend far beyond a purely astronomical context into other domains of science and society. Observational contributions relevant to both questions can be made by measuring minute wavelength shifts of spectral features in astronomical objects. For instance, an Earth-like planet, too faint for a direct observation, can reveal its presence by periodically modifying the radial velocity of its host star and hence Doppler-shifting characteristic features in the stellar spectrum \cite{Mayor1995, Lovis2006}. Similarly, it has been suggested that the changing expansion rate of the Universe could be directly measured by observing the cosmological redshift in distant quasars \cite{Murphy2007, Liske2008}. The major challenge for such measurements is the requirement of a precisely and accurately calibrated astronomical spectrometer capable of detecting frequency shifts equivalent to radial velocities of the order of 10 cm/s or smaller. Conventional approaches of spectrometer calibration typically rely on the emission lines of hollow-cathode gas lamps that are used as calibration markers. However, the limited stability over time, the sparsity and different intensities of emission lines as well as the sensitivity to line blending impose limitations that are incompatible with the observational requirements. Over the last decade it has been realized that laser frequency combs (LFCs) \cite{Telle1999, Jones2000, Udem2002, Cundiff2003, Ycas2012, Glenday2015, McCracken2017} provide new means of wavelength calibration with unprecedented accuracy and precision \cite{Steinmetz2008, Li2008, Wilken2012, McCracken2017a}. Such LFCs are typically derived from mode-locked lasers and consist of large sets of laser lines whose optical frequencies $\rm{\nu_n}$ are equidistantly spaced: $\rm{nu_n = n*f_{rep} + f_0}$ (n is an integer number). The two parameters  $\rm{f_{rep}}$ and  $\rm{f_0}$   are radio-frequencies (RF) accessible by conventional electronics and refer to the pulse repetition rate and carrier-envelope offset frequency of the mode-locked laser. 

	\begin{figure*}[]
	\centering
	\includegraphics[width=0.9\textwidth]{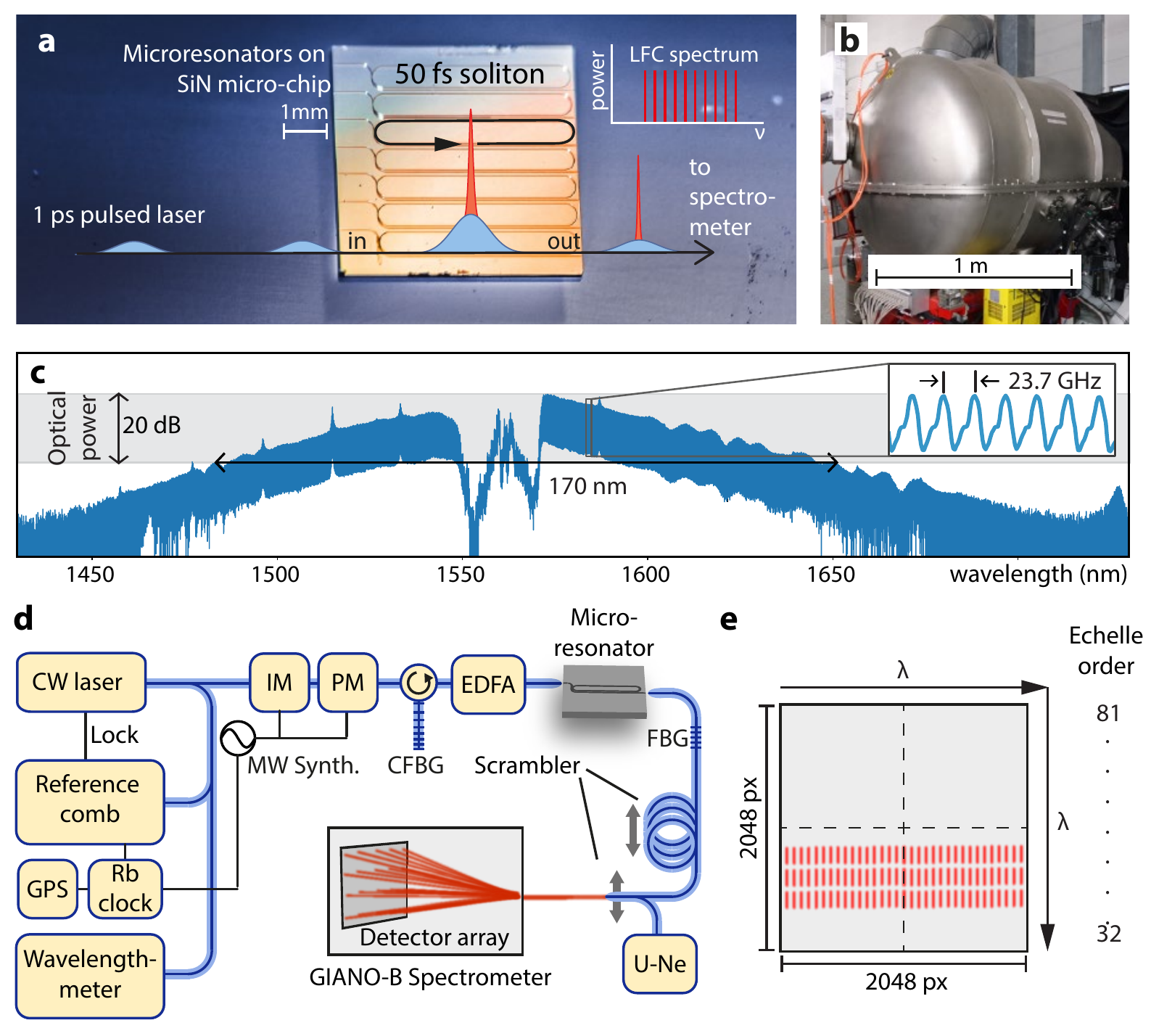}
	\caption{\small \textbf{Microresonator frequency comb generation and setup for astronomical spectrometer calibration. } \textbf{(a)} Silicon-nitride micro-chip with microresonators and scheme of laser frequency comb (LFC) generation in a microresonator by driving with a pulsed laser. The formation of a circulating approx. 50 fs ultra-short soliton pulse inside the microresonator gives rise to a broadband LFC spectrum. The resonator used in this work is made of silicon-nitride has a free-spectral range of 23.7 GHz and a resonance-width of 300 MHz \textbf{(b)} Vacuum chamber of the GIANO-B high-resolution near-infrared spectrometer. \textbf{(c)} Optical spectrum of the microresonator LFC generated with a mode spacing of 23.7 GHz. The useful 20 dB power envelope extends across approximately 170 nm. The ‘spectral hole’ in the center results from a spectral filter introduced to suppress residual driving light. \textbf{(d)} Scheme of the experimental setup. A continuous-wave (CW) laser is intensity and phase modulated (IM/PM) at the microwave frequency of 11.85 GHz, compressed into picosecond pulses by a chirped fiber-Bragg grating and amplified by an erbium-doped fiber amplifier (EDFA). The light is coupled to the microresonator chip where soliton formation gives rise to a broadband optical frequency comb. A fiber-Bragg grating (FBG) is used to remove residual driving light. Before coupling to the spectrometer the comb is transferred from a single-mode into a multi-mode fiber and two spatial mode-scramblers are used to avoid modal noise (speckles). In the GIANO-B spectrometer the light is cross-dispersed and projected onto a 2-dimensional detector array. Alternatively to the LFC, light from a U-Ne hollow-cathode lamp can be injected into the spectrometer. In order to provide absolute wavelength calibration, the CW laser is locked to a self-referenced 100 MHz repetition-rate mode-locked laser. A wavelength-meter provides an approximate measurement of the CW laser’s wavelength such that the lock to the mode-locked laser can be used for an exact and unambiguous determination. Both the microwave synthesizer driving the modulation of the CW laser and the reference comb are referenced to a 10 MHz signal provided by a GPS-disciplined rubidium atomic clock. \textbf{(e)} Scheme of the detector array with 2048 x 2048 pixels. }
	\label{fig1}
	\end{figure*}

Via self-referencing and stabilization schemes,  $\rm{f_{rep}}$ and  $\rm{f_0}$  can be measured and locked to an atomic RF standard leading to a stable optical spectrum with exactly known optical frequencies. While in principle such stabilized LFC spectra would be ideal for astronomical spectrometer calibration, the laser’s repetition rate $\rm{f_{rep}}$, and hence the line spacing, is typically well below the required mode spacing of at least 10 GHz (in order to be resolvable by an astronomical spectrometer). Therefore, previous work used several stages of actively stabilized Fabry-P\'erot filtering cavities in order to thin out the comb spectrum. Besides high experimental complexity, imperfect filtering and nonlinear optical effects effectively shift the apparent frequency of the transmitted mode and constitute sources of systematic error \cite{Chang2010, Chang2012, Probst2013}. Another approach is frequency combs derived from electro-optically modulated continuous-wave (CW) lasers \cite{Yi2016, Beha2017, Torres-Company2014, Carlson2017}. When modulating with microwave frequencies $>$ 10 GHz, no mode filtering is required. However, if not suppressed via a stabilized optical filter cavity, the phase noise inherent to the microwave modulation source can lead to severe linewidth broadening in the wings of very broadband LFC spectra, introducing another source of unwanted line shifts and systematic error.

	\begin{figure*}[]
	\centering
	\includegraphics[width=0.9\textwidth]{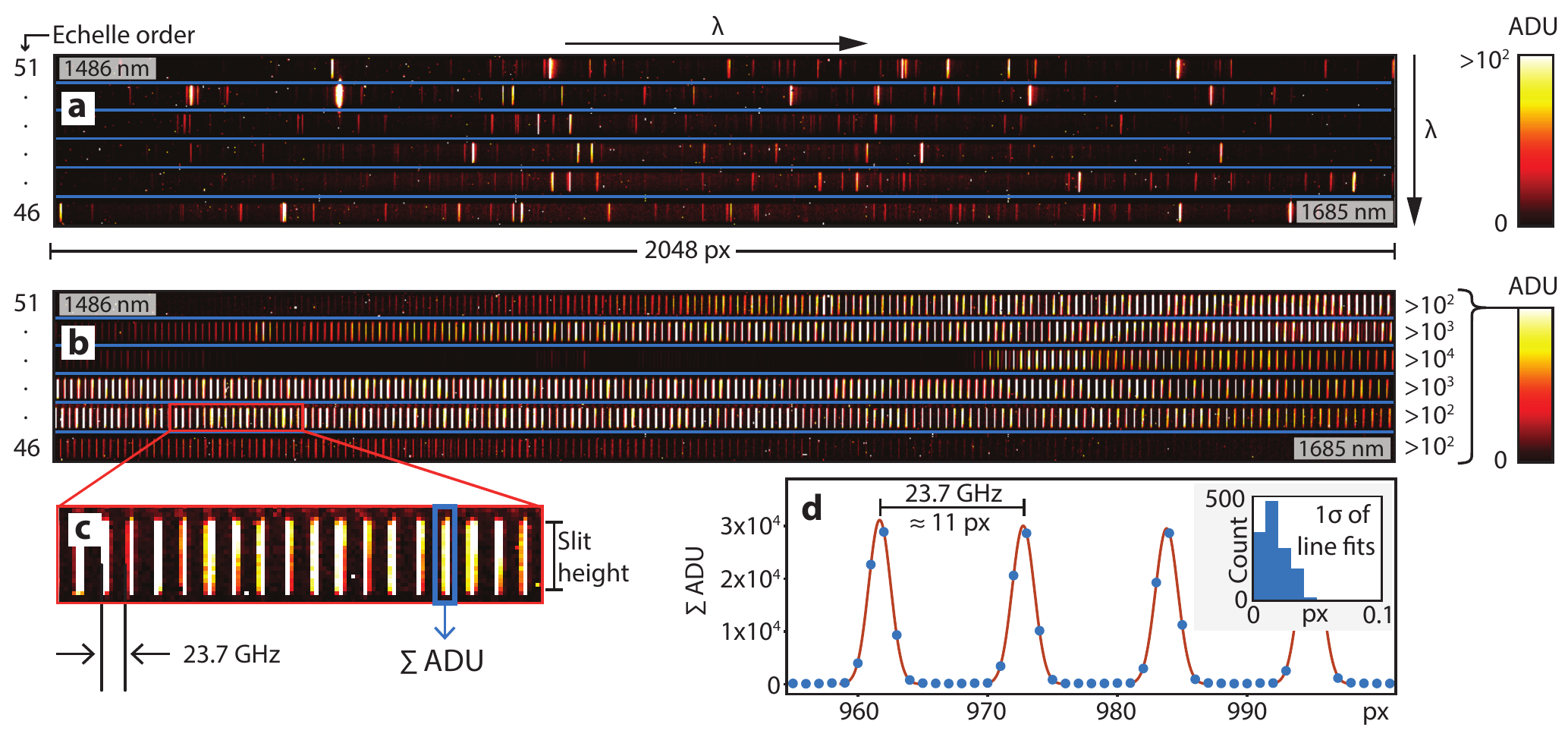}
	\caption{\small \textbf{Spectrometer data: } \textbf{(a)} Calibration spectrum of a U-Ne hollow-cathode lamp recorded in 120 s of exposure time. The graph shows 6 Echelle-orders covering the wavelength range from 1486 to 1685 nm. The wavelength increases from left to right and from top to bottom. The optical intensity is measured in analog-to-digital units (ADU). \textbf{(b)} Spectrum of the microresonator laser frequency comb recorded in 10 s of exposure time. For better visibility each order has a different color scale (cf. scale-bar). The absence of calibration lines in parts of Echelle-order 49 results from the fiber-Bragg grating as explained in the main text. \textbf{(c)} Zoom into panel (b) showing well-resolved comb lines spaced by 23.7 GHz. To enhance the signal-to-noise ratio ADUs can be summed along the vertical direction (spectrometer slit). \textbf{(d)} Summing of ADU results in the one dimensional data along the horizontal pixel-axis. Gaussian line-fitting is used to determine the exact pixel position of the laser frequency comb line. The inset shows the histogram of 1 $\rm{\sigma}$ uncertainties for the line-fits.}
	\label{fig2}
	\end{figure*}

Here, unlike all previous approaches, we use a frequency comb that is generated in a Kerr-nonlinear optical microresonator \cite{DelHaye2007, Savchenkov2008, Kippenberg2011}. When driven by a laser Kerr-nonlinear optical frequency conversion can efficiently occur in such microresonators and result in broadband frequency comb spectra. As the resulting line-spacing is inversely proportional to the size of the resonator, microresonators naturally give rise to a LFC with a suitable line-spacing in excess of 10 GHz without spectral filtering. Narrow linewidth and a smooth spectral envelope are important properties of a LFC for spectrometer calibration. In this regard, the recent discovery of temporal dissipative Kerr-soliton (DKS) in microresonators \cite{Leo2010, Herr2014, Suh2016, Grudinin2017, Cole2017, Joshi2016, Wang2016, Pavlov2017, Brasch2015, Obrzud2017, Weiner2017} has opened entirely new possibilities, which are explored here for the first time outside the protected environment of an optics laboratory and applied in the challenging application of spectrometer calibration. DKS can be generated both in CW and synchronously pulse-driven microresonators. In both cases, an ultra-short soliton pulse is formed inside the resonator, which gives rise to a coherent and broadband optical spectrum. Driving laser noise outside the microresonator’s bandwidth cannot enter the resonator and can hence not affect the comb generation. In this work, we use a ring-type microresonator (Figure ~\ref{fig1}a) fabricated on a silicon-nitride chip \cite{Moss2013} and characterized by a free-spectral range (FSR) of 23.7 GHz and a linewidth of 300 MHz. As it enables direct control over the soliton’s repetition rate and carrier-envelope offset frequency \cite{Obrzud2017, Jang2015}, we extend the pulsed laser driving scheme demonstrated in Fabry-P\'erot type microresonators to the present ring-resonator geometry. The driving laser source is a 1560 nm CW laser (approx. 1 W on the chip) that is electro-optically modulated at 11.85 GHz (half the FSR) and compressed into pulses of pico-second duration. This novel configuration, where the soliton is only driven every other roundtrip \cite{Obrzud2018}, is possible as the soliton roundtrip time is much shorter than the resonator’s decay time. Effectively the soliton adopts twice the modulation frequency as its repetition rate. The resulting spectrum is shown in Figure ~\ref{fig1}c with an underlying soliton pulse of 50 fs duration (100x shorter than the driving laser pulses).  During operation the microresonator self-locks to the driving laser \cite{Carmon2004} so that no active feedback-loop is required. In order to provide absolute frequency calibration, the CW laser frequency (which determines the microresonator comb’s offset frequency) is linked to the 10 MHz RF signal of a GPS-disciplined rubidium atomic clock via a self-referenced 100 MHz repetition rate mode-locked laser as detailed in Figure ~\ref{fig1}d. A fiber-Bragg grating (FBG) was added after the microresonator to suppress the residual driving laser light. The resulting spectral hole in the center of the frequency comb can be avoided in future work by using a critically coupled microresonator, an adapted FBG filter or a drop-port configuration \cite{Wang2016, Obrzud2017}.

	\begin{figure*}[]
	\centering
	\includegraphics[width=0.9\textwidth]{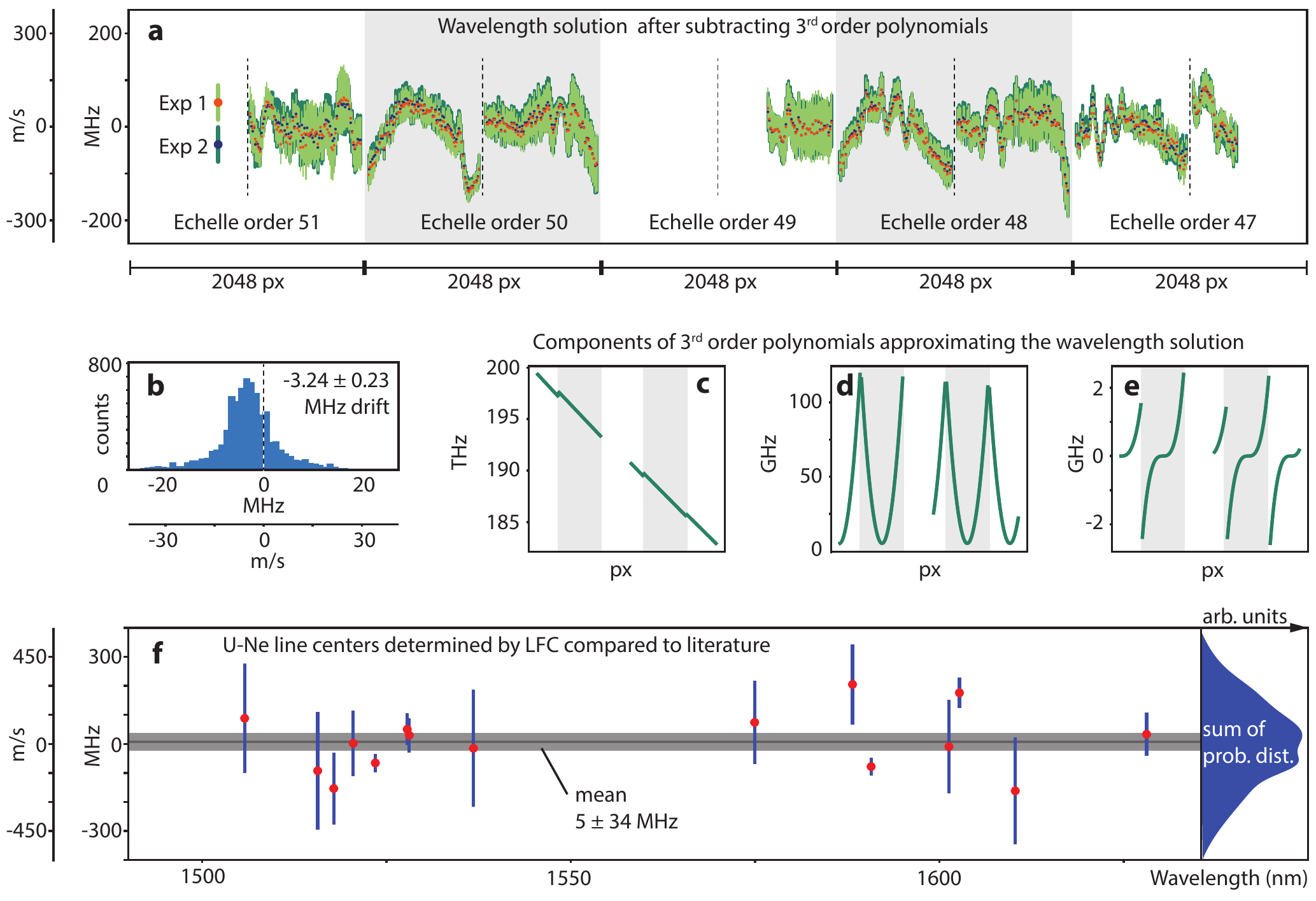}
	\caption{\small \textbf{Results of microresonator based calibration.} \textbf{(a)} Calibration data for two LFC exposures (“Exp 1”, “Exp 2”) in units of optical frequency and radial velocity at 1.5 $\rm{\mu m}$ wavelength derived from two LFC exposures separated by 5 minutes and with 10 s exposure time each. For each Echelle-order a 3$\rm{^{rd}}$ order polynomial fit has been subtracted in order to make subtle features visible to the eye (polynomial detailed in panel c, d, e). Each data point corresponds to a detected LFC line and the errorbars indicate the 1 σ uncertainty in determining the line’s pixel position. The dashed vertical lines indicate the center of each order, where a discontinuity can be observed due to detector stitching. (b) Pixel-by-pixel difference between the two wavelength solutions in (a). A global drift of the spectrometer by 3.24 $\pm$ 0.23 MHz is visible. (c, d, e) First, second and third order components of the polynomial fits that were subtracted from the LFC based wavelength solution to yield the trace shown in (a). (f) Comparison of individual U-Ne line frequencies determined by the LFC calibration with line frequencies in literature. The differences of those values are shown as a function of their wavelength (1 $\rm{\sigma}$ uncertainty intervals result from the least-squares fitting of U-Ne lines). The mean of all differences is compatible with zero. The ‘sum of probability distributions’ (on the right) illustrates the symmetry of the difference values around zero. }
	\label{fig3}
	\end{figure*}

The demonstration of microresonator LFC-based wavelength calibration is performed on the GIANO-B high-resolution near-infrared spectrometer at the Telecopio Nazionale Galileo (TNG) on La Palma, Spain \cite{Oliva2012} (Figure ~\ref{fig1}b). As a cross-dispersing Echelle-type spectrometer GIANO-B projects the near-infrared spectrum from 0.9-2.4 $\rm{\mu m}$ onto a 2-dimensional detector array of 2048 x 2048 pixels (Hawaii2RG). On the detector, the spectrum is organized in 50 Echelle-orders, covering consecutive wavelength intervals. The wavelengths contained in each Echelle-order are dispersed horizontally across the detector and the individual Echelle orders are separated vertically (where increasing order number implies shorter wavelength). The extent of a single optical frequency on the detector is given horizontally by the spectrometer’s point-spread-function (PSF) and vertically by the height of the spectrometer’s entrance slit. Prior to injecting the LFC light into the spectrometer two scramblers based on mechanically actuated multi-mode fibers \cite{Baudrand2001} were used in order to avoid modal noise (speckles) on the detector (cf. Figure ~\ref{fig1}d). The goal of the wavelength calibration is to provide an accurate and precise pixel-to-wavelength mapping, which is also referred to as “wavelength-solution”.

Figure ~\ref{fig2}a shows the detector image of the conventional calibration light source, i.e. a uranium-neon (U-Ne) hollow-cathode lamp after a 120 s exposure. The sparsity of U-Ne emission lines is a major limitation in deriving the wavelength solution. In contrast, the microresonator LFC (cf. Figure ~\ref{fig2}b) provides a much denser grid of well-resolved calibration lines. The observed LFC line-shapes are given by the spectrometer’s PSF, which is large compared to the width of the LFC’s modes. In a first step towards the wavelength solution, the signal (measured in analog-to-digital units, ADU) in the inner 20 pixel (along the vertical direction) of each detected LFC line is summed in order to increase the signal-to-noise ratio (cf. Figure ~\ref{fig2}c). In the resulting 1-dimensional data (cf. Figure ~\ref{fig2}d) the horizontal pixel position of each LFC line is determined via least-squares fitting of a Gaussian PSF-model in a 5-pixel-wide window around each pixel with a peak ADU count. The 1 σ uncertainty of a single line fit is typically well below 5\% of a pixel’s width corresponding to less than 100 MHz of fit uncertainty (cf. Figure ~\ref{fig2}d, inset).  This uncertainty is related to fundamental photon-noise \cite{Bouchy2001}, in the ADU counts. LFC lines in the vicinity of hot pixels or cosmic ray induced artifacts are masked and excluded from the analysis. Next, each detected LFC line pixel-position is assigned an exact optical frequency based on the known parameters of the LFC. The results of two LFC exposures (comb line frequency vs. pixel-position) observed with 5 minutes of separation in time are shown in Figure ~\ref{fig3}a. Here, for each Echelle-order a 3$\rm{^{rd}}$ order polynomial fit was subtracted in order to make subtle details in the calibration data visible (The components of the 3rd order polynomial are shown in Figure ~\ref{fig3}c,d,e). In order to reduce the effect of line-fitting uncertainty a 5-point sliding box average is applied to this calibration data. This is justified as the uncertainties of the calibration points are significantly larger than the underlying structures in the frequency-versus-pixel data. The two LFC sets of calibration data exhibit the same characteristic features, which also repeat with high similarity for each Echelle order. Notably, discontinuities in calibration data are visible in the center of each Echelle order (Figure ~\ref{fig3}a, dashed lines). This is a result of the detector being comprised of four 1024 x 1024 pixel arrays that are arranged next to each other to form the full array. Here, the slightly different distances between neighboring pixels at the arrays’ interfaces imply a discontinuous frequency-versus-pixel behavior. In a final step, the actual pixel-to-wavelength mapping (wavelength solution) can be found for any pixel position by interpolating between the LFC calibration data points.

In order to quantify the quality of the LFC-based wavelength solution, a conservative upper limit of the global precision can be found by taking the pixel-wise difference between the two wavelength-solutions. This yields the difference histogram shown in Figure ~\ref{fig3}b, which reveals a global spectrometer drift of 3.24 $\pm$0.23 MHz (4.86 $\pm$ 0.35 m/s). The single wavelength solution precision is hence 25 cm/s, in agreement with an estimate of the fundamental photon noise of approximately 20 cm/s \cite{Bouchy2001}. Despite being a first demonstration, the achieved calibration is already significantly superior to the conventional calibration approaches and of immediate relevance to advanced astronomical observation. Finally, we verify the microresonator LFC’s consistency with the conventional U-Ne standard. To this end, a U-Ne spectrum that was taken in-between the two LFC exposures is analyzed. Careful inspection of the U-Ne spectrum led to the selection of 15 emission lines that were well isolated and free of artifacts such as line blending or modulated background signals. An average between the two LFC wavelength solutions is used to attribute optical frequencies to the U-Ne lines, whose pixel position were found again via least-squares Gaussian-PSF fitting. Subtraction of literature values \cite{Oliva2012, Redman2011} from the LFC-based frequency values results in a mean difference consistent with zero (cf. Figure ~\ref{fig3}f). This proves the consistency of the LFC based calibration with the established calibration standard.

In summary, we have shown that microresonators provide novel and natural means of generating widely-spaced optical frequency combs suitable for astronomical spectrometer calibration. Already in this first demonstration, we obtain a wavelength calibration of the order of 10 cm/s, which can be further improved by increasing the LFC’s spectral span. Importantly, this is not a technological limitation and indeed microresonator spectra spanning the entire near-infrared regime have already been achieved \cite{Brasch2015, Del'Haye2011, Okawachi2011, Li2016,Pfeiffer2017}. In addition, efforts of extending microresonator frequency combs into the visible wavelength regime are ongoing \cite{Xue2015, Lee2017}. As such microresonator frequency combs hold great potential as calibration sources for the next-generation of astronomical instruments in planet-hunting and cosmological research.
\\
\\
{\scriptsize
This work was supported by the Swiss National Science Foundation (grants 200021-166108 and 163387), the NCCR-PlanetS and NCCR-PlanetS 
Technology Platform, the NCCR-QSIT (51NF40-160591), the Canton of Neuchatel and INAF Progetto Premiale WOW.}


\bibliographystyle{unsrt}

\end{document}